# Superconductivity in topologically nontrivial material $Au_2Pb$

Ying Xing[1,2,#], He Wang[1,2,#], Chao-Kai Li[1,2,#], Xiao Zhang[1,2], Jun Liu[3], Yangwei Zhang[1,2], Jiawei Luo[1,2], Ziqiao Wang[1,2], Yong Wang[3], Langsheng Ling[4], Mingliang Tian[4], Shuang Jia[1,2], Ji Feng[1,2,*], Xiong-Jun Liu[1,2], Jian Wei[1,2,*], Jian Wang[1,2,*]

[1]International Center for Quantum Materials, School of Physics, Peking University, Beijing 100871, China

[2]Collaborative Innovation Center of Quantum Matter, Beijing 100871, China

[3]Center of Electron Microscopy, State Key Laboratory of Silicon Materials, Department of Materials Science and Engineering, Zhejiang University, Hangzhou 310027, China

[4]High Magnetic Field Laboratory, Chinese Academy of Sciences, Hefei 230031, China

[#]Authors contributed equally to this work.

Correspondence and requests for materials should be addressed to

jianwangphysics@pku.edu.cn (J. Wang); weijian6791@pku.edu.cn (J. Wei);

jfeng11@pku.edu.cn (J. Feng)

# Abstract

The search for nontrivial superconductivity in novel quantum materials is currently a most attractive topic in condensed matter physics and material science. The experimental studies have progressed quickly over the past couple of years. In this article, we report systematic studies of superconductivity in $Au_2Pb$ single crystals. The bulk superconductivity (onset transition temperature, $T_c^{onset}$= 1.3 K) of $Au_2Pb$ is characterized by both transport and diamagnetic measurements, where the upper critical field $H_{c2}$ shows unusual quasi-linear temperature dependence. The superconducting gap is revealed by point contact measurement with gold tip. However, when using tungsten (W) tip, which is much harder, the superconducting gap probed is largely enhanced as demonstrated by the increases of both $T_c^{onset}$ and upper critical field ($H_{c2}$). This can be interpreted as a result of increase in density of states under external anisotropic stress imposed by the tip, as revealed by first-principles calculations. Furthermore, novel phase winding of the pseudospin texture along k-space loops around the Fermi energy is uncovered from the calculations, indicating that the observed superconductivity in $Au_2Pb$ may have nontrivial topology.

**Introduction**

The recent discovery of three-dimensional (3D) Dirac semimetals[1-12], as an intermediate state between a trivial insulator and a topological insulator, has stimulated extensive research on these materials. In 3D Dirac semimetals, the conduction and valence bands contact only at Dirac points in the Brillouin zone, and gap formation is forbidden by crystalline symmetry. Much recent excitement has surrounded the materials like $Na_3Bi$[2] and $Cd_3As_2$[3], in which the Dirac cone semimetal states were predicted theoretically and soon identified by Angle resolved photoemission spectroscopy (ARPES) experiments.[4, 6, 11] Surprisingly, the superconductivity induced by hard point contact on $Cd_3As_2$ was recently reported and unconventional superconducting order parameter was suggested by the zero-bias conductance peak (ZBCP) and double conductance peaks (DCPs) symmetric around zero bias.[13,14] Additionally, pressure induced superconductivity in $Cd_3As_2$ was further confirmed by hydrostatic pressure experiment.[15]

Another promising material is the cubic Laves phase $Au_2Pb$, which reveals the signature for the symmetry-protected Dirac semimetal state at temperature above 100 K. However, after the structural phase transition the Dirac cone is gapped below 100 K.[16] Here we report systematic studies of low temperature transport and diamagnetic properties, point contact (PC) measurements, and first-principles calculations of $Au_2Pb$ single crystal. The superconductivity of $Au_2Pb$ is characterized, and the anomalous quasi-linear $H_{c2}$(T) in $Au_2Pb$ suggests the presence of an unconventional superconducting state. The $T_c^{onset}$ and $H_{c2}$ revealed by the PC measurement with a gold tip is consistent with that by the bulk transport and magnetization measurements, but with a hard W tip, the gap is largely enhanced as shown by the increases of both $T_c^{onset}$ and $H_{c2}$. By comparing the computed Fermi surfaces of $Au_2Pb$ of relaxed lattice structure and those under 1% uniaxial compression, we find that the Fermi surfaces are enlarged and the density of states near the Fermi energy increases under the applied external pressure, which agrees with the observed increase of $T_c^{onset}$ in the PC measurements. More importantly, the possibility of topological superconductivity is suggested by first-principles calculations, which show nontrivial topological properties of the projected pseudospin texture corresponding to the p- and d-orbitals near the Fermi energy, as well as the experimental observations of anomalous quasi-linear $H_{c2}(T)$ behavior and unusual PCS feature detected by W tip. These characteristics make $Au_2Pb$ a potential candidate material for topological superconductor.

**Results**

**Sample Structure.** High-quality single crystal of $Au_2Pb$ samples were synthesized by self-flux method in the evacuated quartz ampoule (see Methods). The powder X-ray diffraction pattern is given in Fig. 1a, indicating a crystal structure of $Au_2Pb$ and exhibiting a cubic Laves phase with $a_1=a_2=a_3=7.9603$Å at room temperature. The single crystal X-ray diffraction pattern only shows (111) reflections, suggesting the sample surface of the crystal is (111) plane (inset of Fig. 1a). The atomic high resolution transmission electron microscopy (HRTEM) image (Figure 1b), together with the selected area electron diffraction pattern (Figure 1c), further demonstrate the high quality single crystal nature of $Au_2Pb$ samples.

**Superconductivity.** The superconductivity in $Au_2Pb$ has been confirmed by our systematic low temperature transport and diamagnetic measurements. Figure 2a displays the sample resistivity ($\rho$) as a function of temperature at zero magnetic field. The behavior is metallic with an anomaly of $\rho(T)$ at 100 K due to a structural phase transition (cubic Laves phase above 100 K)[16]. A steep step at 50 K and hysteresis at 40 K are observed with cooling and heating measurements, owing to structural transition to orthorhombic phase below 40 K (the inset of Fig. 2a). Upon cooling down to lower temperature, bulk $T_c^{onset}$ around 1.3 K and zero resistance transition temperature $T_c^{zero}$ around 1.18 K are observed, comparable to the previous reports[16, 17]. The superconductivity in $Au_2Pb$ is further confirmed by Meissner effect measurements. In Fig. 2b, the susceptibility ($\chi$) versus temperature curves show quite sharp drops at 1.15 K, in good agreement with $T_c^{zero}$ from $\rho(T)$ curve. Magnetic field dependence of the magnetization ($M(H)$) curves at various temperatures(inset of Fig. 2b) exhibit the expected quasi-linear behavior at low fields but deviate from linearity above the lower critical field $H_{c1}$. Figure 2c reveals the suppression of the superconducting state of $Au_2Pb$ by the magnetic field perpendicular to (111) plane ($H_\perp$). The superconducting transition becomes broader and shifts to lower temperature with increasing fields. Magnetotransport measurements were carried out at various temperatures from 0.18 K to 1.4 K (Fig. 2d). The onset $H_{c2}$, defined as the field above which the $Au_2Pb$ sample becomes the normal state, is shown in Fig. S1. $H_{c2}$ linearly increases with decreasing temperature down to $T_c/6$, yields $dH_{c2}(T)/dT|_{T=T_c} \approx 0.054$ T/K near $T_c$. We calculate the reduced critical field $h^*=H_{c2}/T_c dH_{c2}/dT|_{T=T_c}$ (inset of Fig. 2d) to compare the data to the known models for s-wave superconductors (Werthamer–Helfand–Hohenberg theory, WHH, $H_{c2} \approx 0.7T_c dH_{c2}/dT|_{T=T_c}$, $h^*(0) \approx 0.7$)[18] and spin-triplet p-wave superconductors ($h^*(0) \approx 0.8$)[19]. Obviously, the $h^*$ relation is close in the form to that of a polar p-wave state, which is suggestive of the finite triplet contribution to the pairing state in $Au_2Pb$.[19,20] The upper critical field $H_{c2}(T)$ shows little anisotropic property for perpendicular & parallel magnetic field (Fig. S2). Furthermore, we plotted the normalized magnetoresistivity (MR = $(\rho(H)-\rho(0))/\rho(0)\times100\%$) of $Au_2Pb$ as the field up to 15 T, from 1 K to 200 K (see Fig. 2e). In contrast to the classical quadratic MR in metals and semiconductors, the sample presents non-saturating linear-like MR at lower temperatures and higher magnetic fields. With increasing temperatures, the exponent $\alpha$ (MR $\propto H^{\alpha}$) varies between 1 and 2 and exhibits anomalies when temperatures undergoing structural phase transitions around 40 K, 100 K (Fig. S3b). In the parallel field ($H$ // (111) plane: $H_{//}$) configuration, Figure 2f demonstrates how the MR behavior changes when the direction of the magnetic field is rotated from 0° ($H$ // [$\bar{1}\bar{1}$2] $\perp I$) to 90° ($H$ // [1$\bar{1}$0] // $I$) at 2.5 K. Classically, the resistance has no respond to the applied external magnetic field parallel to the excitation current. While in our situation, MR is quasi-linear-dependent on magnetic field (MR $\propto H^{1.09\pm0.02}$, inset of Fig. 2f). The linear magnetoresistance behavior has usually been observed in semiconductor[21], semimetals[22], topological insulators[23-25] and Dirac/Weyl semimetals[10, 26, 27]. One general interpretation is inhomogeneity in materials, but the inhomogeneity does not seem to play an important role here, since our $Au_2Pb$ crystals show good single crystal quality. Alternatively, it is also tempting to ascribe the linear magnetoresistance to Abrikosov's quantum magnetoresistance[28], however, the estimated carrier density $10^{22}$ cm$^{-3}$ (see Fig. S4) is too high for this model to be applicable over entire field range. It is noticed that this longitudinal linear MR ($H$ // $I$) due to the balanced hole and electron populations, was ever observed in type II Weyl semimetal candidate $WTe_2$.[22,29] The linearity here may originate from the multiband nature of $Au_2Pb$.

**$T_c^{onset}$ and $H_{c2}$ enhancements in PC measurement with a W tip.** PC Andreev reflection spectroscopy is a powerful tool to probe the order parameters of superconductor[30]. The temperature dependence of zero bias differential resistance of PC between a gold tip and $Au_2Pb$ single crystal is shown in Fig. 3a. The presence of a significant resistance drop at the onset temperature $T_c^{onset}$ = 1.13 K at zero magnetic field and the absence of this drop at 0.03 T applied perpendicular to the (111) surface indicate a superconducting transition, which is similar to that in the bulk measured by standard four-electrode method (see Fig. 2). However, when measured with a W tip as shown in Fig. 3b, the onset temperature increases to 2.1 K, higher than the bulk value. And the perpendicular field ($H_\perp$) used to fully suppress the superconductivity transition is 0.755 T, also much higher than the bulk value. The point contact spectra (PCS) at different temperatures with a gold tip and a W tip are shown in Fig. 3c and Fig. 3d respectively. For the PC with a gold tip, DCPs are observed at 0.5 K and gradually get smeared with increasing temperature. The resulted broad conductance peak totally diminishes at 1.1 K, consistent with the transition temperature in d$V$/d$I$($T$) as shown in Fig. 3a. For the PC with a W tip, similar DCPs feature is clearly shown at lower temperatures, and the broad conductance enhancement totally diminishes at $T$ = 2.1 K, consistent with $T_c^{onset}$ in Fig. 3b. We note that besides the DCPs feature, there are also conductance dips outside the conductance enhancement regime as shown in Fig. 3d for temperatures higher than 1 K, which evolve to broad valleys below 1 K (around bulk $T_c$ shown in Fig. 2b). These conductance dips might be related to the critical current effect[31]. Alternatively, conductance dips at around the gap energy with a broad zero bias conductance peak could be resulted by the helical p-wave order parameter[32]. Another unusual feature of the PCS is a small splitting of the conductance peaks at 0.3 K, which can be fitted by the modified Blonder-Tinkham-Klapwijk (BTK) model[33,34] for two gaps, while the double conductance dips and broad conductance valleys cannot be fitted by conventional s-wave pairing symmetry considerations (Figure S5).

The magnetic field dependence of PCS at 0.5 K is shown in Fig. 3e and Fig. 3f for PC with a gold tip and a W tip respectively. All the PCS show gradual suppression of the superconducting features with increasing magnetic field. The $H_{c2}$ for PC with a gold tip is about 0.02 T at 0.5 K, which is close to the bulk value (~ 0.05 T) obtained by standard four-electrode measurements, while the $H_{c2}$ value of PC with a W tip is about 0.60 T, which is more than ten times larger than the bulk value.

**Analysis of point contact Results:** For PC with a W tip (harder than gold tip), both $T_c^{onset}$ and $H_{c2}$ are larger than bulk values (Figure S6) .The $T_c$ enhancement seems related to the pressure applied by the hard W tips, which may cause changes of the band structure or lifting of degeneracy of multiple order parameters[35, 36]. The conventional cause of $H_{c2}$ enhancement in PC is the reduction of the mean free path $l$ in the superconductor as a result of structural defects and/or impurities at interface[37, 38], which leads to a decrease of the coherence length and increase of $H_{c2}$. However, this cannot explain the concomitant enhancement of $T_c$ and $H_{c2}$ in our situation. In addition, similar superconductivity enhancement for PC with PtIr tips on $Au_2Pb$ samples was observed, where PtIr tips are also much harder than gold tips. Both the PC results of W tip and PtIr tip suggest that the superconductivity in $Au_2Pb$ is sensitive to pressure and/or possible doping effect. Magnetoresistance measurements with field applied parallel or perpendicular to sample surface show little anisotropy, which rules out the possibility of surface superconductivity (Figure S2).

**The first-principles calculations.** Two important clues in the experimental data are worth remarking. First, for the point contact spectra with a W tip, the double conductance dips feature cannot be adequately described by the BTK theory with s-wave pairing symmetry, and points to possible unconventional superconductivity of this material. Second, for transport experimental results, the quasi-linear $H_{c2}(T)$ data imply the expectation for a p-wave state and deviate significantly from the WHH theory for an s-wave superconductor. Therefore, it is important to analyze the electronic structure of $Au_2Pb$ in detail to identify possible unusual topological features.

In order to examine the effects of point-contact tip indentation on the sample's local electronic structure, we compare the computed Fermi surfaces of $Au_2Pb$ of relaxed structure and those under 1% uniaxial strain along the [111] direction. Although the Fermi surfaces are rather complex and appear to be composed of multiple sectors throughout the Brillouin zone (see Fig. S7), it is evident that the Fermi surfaces are enlarged under the applied external pressure. The augmentation of Fermi surfaces and the accompanying increase in the density of states at the Fermi energy (shown in Fig. 4d) favor higher superconducting $T_c$, which agrees with the observed increase of $T_c$ in the PC measurements. It should also be remarked that the size of Fermi surfaces near the Brillouin zone path Γ-R are large and noticeably responsive to external pressure, as shown in Fig. 4a-c. Thus, in the following we will focus on the regions near this sector (referred to as the T-sectors hereafter) of the Fermi surfaces.

It is of particular interest to investigate the geometric phase of the bands near the Fermi surfaces. By projecting the Kohn-Sham wavefunctions into various local orbital basis (Figure S8), we find that the dominant band components near the Fermi surfaces are parity-even d orbitals of Au and parity-odd p orbitals of Pb. We choose two closed paths on the valence band around one of the T-sectors, and calculate topological invariants through the phase difference, $\phi(k) = \arg\langle d | k \rangle - \arg\langle p | k \rangle$, of the projection coefficients of Kohn-Sham wavefunctions of the highest occupied onto the Au-d and Pb-p orbitals along these loops. The two loops are carefully chosen to avoid band degeneracy at the band crossing [see the dark blue curve in Fig. 5a]. In Fig. 5b, a unit vector $n = [\cos\varphi, \sin\varphi]$, is plotted for each k-point on the loop to visualize the orbital texture.

A few interesting features are observed for the projected orbital textures. In particular, the winding numbers are zero for the orbital textures on the inner loop [see e.g. a typical orbital texture on the inner loop in Fig. 5b]. In contrast, for the outer loop, the orbital texture corresponding to various Au-d and Pb-$p_x$ orbitals show a nontrivial topology, with the winding number equals -1. On the other hand, from Fig. 6 we find that an "8" shape projected nodal line is numerically confirmed on a certain Brillouin zone slice, while on another Brillouin zone slice a circle shape projected nodal line is observed. These intriguing observations hint at nontrivial band structure and topological properties of the T-sectors.

To capture the main features of the band structure around the T-sectors, we introduce an *ad hoc* two-band effective Hamiltonian to describe the physics regarding the Au-d and Pb-$p_x$ orbitals:

$$H_{eff} = M(\mathbf{k})\mathbf{k}\cdot\sigma, \qquad (1)$$

Where k denotes the momentum relative to the central point of the "8" shape nodal line shown in Fig. 6a, and $M(\mathbf{k})$ is a polynomial of k, taken as $M(\mathbf{k}) = m_1 - \sqrt{m_2^2 + k^2}$ with $m_1 > m_2$. The energy spectra read $E_\pm = \pm|M(\mathbf{k})\mathbf{k}|$, exhibiting a nodal point at $k=0$ and a nodal surface at $k = \sqrt{m_1^2 - m_2^2}$. One can verify that the above Hamiltonian leads to the band structure and orbital texture consistent with the DFT results. Actually, the valence band wavefunction of the Hamiltonian is

$$|u_k\rangle = \begin{bmatrix} -\sin\dfrac{\theta}{2} e^{-i\phi/2} \\ \cos\dfrac{\theta}{2} e^{i\phi/2} \end{bmatrix}, \qquad (2)$$

Where $\theta$ and $\phi$ are the polar and azimuthal angles of *k*, respectively. From the wavefunction we can see that the orbital texture of the Hamiltonian (1) is determined solely by the $\hat{k}\cdot\sigma$ factor. The form of $M(\mathbf{k})$ has been chosen to mimic the spectra from DFT calculation, reproducing the nodal surface near the Fermi level. It is easy to see that the projected orbital texture is dependent only on $\phi$ evolved along each path. The orbital texture on any loop that encloses the z-axis ($k_x = k_y = 0$) connecting the north and south poles of the Bloch sphere, has a winding number 1 or -1. In contrast, if the loop excludes the z axis, the orbital texture is trivial with the winding number being 0. Accordingly, a loop enclosing the nodal line on the $k_z$ = constant plane (i.e. the crossing line between the plane with fixed $k_z$ and the nodal surface) always encloses the z axis, giving a nonzero winding number of the orbital texture [the outer loop of Fig. 5b]. However, a loop inside the nodal line may or may not enclose the z axis. In the latter case the winding number of the orbital texture is zero, which is understood to be the case of the inner loop of Fig. 5b.

**Discussion**

The nontrivial topology of the orbital texture suggests the possibility of topological superconductivity. Note that the whole Brillouin zone includes eight different sectors around Fermi energy, with the other seven equivalent copies of the effective Hamiltonian Eq. (1) obtained by time-reversal and symmetry of the orthogonal space group. In particular, the time-reversal copy of the Hamiltonian (1) can be described by $\Gamma H_{eff} \Gamma^{-1} = M(\mathbf{k})\left[-k_x \sigma_x + k_y \sigma_y - k_z \sigma_z\right]$, where the time-reversal operator $\Gamma = iKs_y$ where $K$ is the complex conjugate operator, and $s_y$ is a Pauli matrix operating on real spin. We note that the Hamiltonian Eq. (1) and its time-reversed copy have opposite spin orientations, which are not explicitly described here. It is readily verified that the topology of the two time-reversed Fermi surfaces is the same. Note that the pairing mechanism cannot annihilate the topological invariants if the Cooper pairs are formed by two Fermi surfaces of the same topology[39, 40]. This implies that the superconductivity due to the pairing between such two time-reversed Fermi surfaces is fully gapped and of nontrivial topology inherited from the single-particle states. Due to the time-reversal symmetry, the resulted 3D topological superconductivity is classified by an integer invariant[41, 42], and thus may be stable in the presence of multiple copies of the above paired Hamiltonians in the Brillouin zone.

The combination of transport, diamagnetic, PC measurements, and the first-principles calculations are used to study $Au_2Pb$ crystals. We characterized the superconductivity of $Au_2Pb$ crystals by transport and diamagnetic experiments. Enhancement of superconductivity is found when it is probed by a hard point contact, and we attribute this to local pressure, which is consistent with the enlarged Fermi surface by theoretical calculations. The anomalous quasi-linear $H_{c2}$ vs. $T$ relation, as well as the gap feature (double conductance peaks with double conductance dips) in PCS, imply possible unconventional superconducting properties in $Au_2Pb$ crystals. Furthermore, the first-principles calculations point to the nontrivial topology of the orbital texture near the dominant Fermi surfaces, which suggests the possibility of topological superconductivity. The results presented in this work indicate the $Au_2Pb$ crystal could be a promising platform for the investigation of topological superconductivity. For full understanding of the superconductivity in $Au_2Pb$, further investigations like thermal transport and ARPES experiments are clearly necessary and interesting.

*Note added*: After the completion of this work, we notice a recent hydrostatic pressure study of bulk $Au_2Pb$[43], where $T_c$ decreases with pressure. It is opposite to our observation. The major difference between two studies is that their pressure is hydrostatic but ours is nearly local uniaxial pressure on (111) crystal face.

**Methods**

**Sample growth and characterization.** Starting materials of high-purity elemental Au and Pb were prepared to synthesize single crystalline $Au_2Pb$, the initial ratio of Au:Pb was 40:60, the extra Pb were used as flux. The materials were sealed in evacuated quartz ampoules, heated to 600 °C, held for 1 day and then slowly cooled down to 300 °C over a period of 30 hours. This was then followed by centrifugation to remove the flux. Before transport and diamagnetic measurements, the samples were etched in a hydrogen peroxide solution for several minutes to remove the residual Pb flux[16].

The data of powder X-ray diffraction (XRD) and single crystal XRD were collected from a Rigaku MiniFlex 600 diffractometer and then refined by a Rietica Rietveld program. Crystal purity, structure and lattice constant ($a_1=a_2=a_3=7.9603$Å) can be retrieved.

To obtain HRTEM images, the $Au_2Pb$ single crystal was examined by a FEI Tecnai G2 F20 S-TWIN TEM operating at 200 kV.

**Resistivity and diamagnetic measurements.** For the transport measurements, the contacts on the single crystals were made by applying the silver paint on the top surface (111) of $Au_2Pb$ samples, with contact resistance less than 1 Ω. The resistance and magnetoresistance were measured in a commercial Physical Property Measurement System (Quantum Design, PPMS-16, d.c. technique), with the Helium-3 option for temperature down to 0.5 K & Dilution option down to 0.05 K. The excitation current of 1 mA was used for the measurements in low temperature regime Angular dependence of magnetoresistance was measured by rotating the sample ((111) surface plane) in a Rotator option based on PPMS, within the instrumental resolution of 0.1°.

For diamagnetic measurement, DC magnetization was studied during the zero-field cooling and field cooling at $H$ = 5 Oe, in a Magnetic Property Measurement System (MPMS 7-XL SQUID) from Quantum Design Company, with a resolution of $10^{-8}$ emu. The magnetic field is applied parallel to the (111) face. The $\chi$ is estimated by DC magnetization.

**The PC measurements.** Our PCS are obtained using a standard lock-in technique in quasi-four probe configuration. PC measurements are realized in the standard "needle-anvil" configuration.

Both W and gold tips are used to make PC on the (111) surface of $Au_2Pb$ single crystals. The W tip is prepared by electrochemical etching method with wire of 0.25 mm diameter, and is hard enough to penetrate through the surface layer and to probe the superconductor underneath. The gold tip is mechanically sharpened from 0.5 mm wire and is relatively soft.

**The first-principles calculations.** The first-principles calculations were performed in the scheme of density-functional theory, as implemented by the VASP package[44]. The projector-augmented wave pseudopotentials were used with Perdew-Burke-Ernzerhof exchange-correlation functional. The energy cutoff of the plane wave basis set was chosen to be 350 eV. The Brillouin zone was sampled by a 13×19×9 grid in the self-consistent calculation. During the relaxation procedure we used a force threshold of 0.01 eV/Å. Spin-orbit coupling was included in the calculation of energy bands and Fermi surfaces, and not included in the calculation of orbital textures in order to avoid the two-fold degeneracy of the bands to obtain a well-defined phase difference.

**Acknowledgements**

**Contributions**

**Competing Interests**

The authors declare no competing financial interests.

**Funding**

This work was financially supported by the National Basic Research Program of China (Grant Nos. 2013CB934600 & 2012CB921300 & 2012CB927400), the Research Fund for the Doctoral Program of Higher Education (RFDP) of China, the Open Project Program of the Pulsed High Magnetic Field Facility (Grant No. PHMFF2015002), Huazhong University of Science and Technology, and the National Natural Science Foundation of China (Grant No. 11474008).

**Figures**

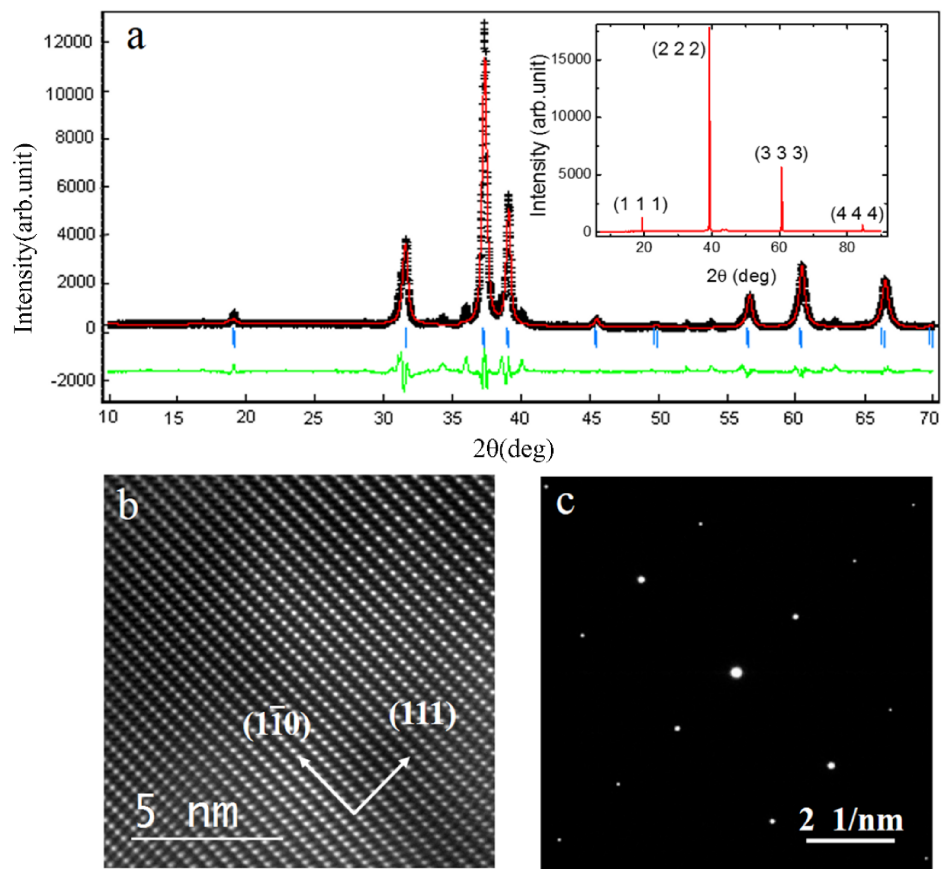

**Figure 1 Sample characterization.** (**a**) X-ray diffraction patterns at room temperature for the Au$_2$Pb samples. The major diffraction peaks can be indexed with a cubic Laves phase of Au$_2$Pb. Inset: The X-ray diffraction pattern from the basal plane crystal surface of Au$_2$Pb only shows (111)$_n$ reflections, which indicates the measured crystal plane of the crystal is (111) plane. (**b**) High-resolution transmission electron microscopy image of Au$_2$Pb crystal. (**c**) Selected area electron diffraction image looking down from the [$\bar{1}\bar{1}2$] zone axis showing the reciprocal lattice of Au$_2$Pb.

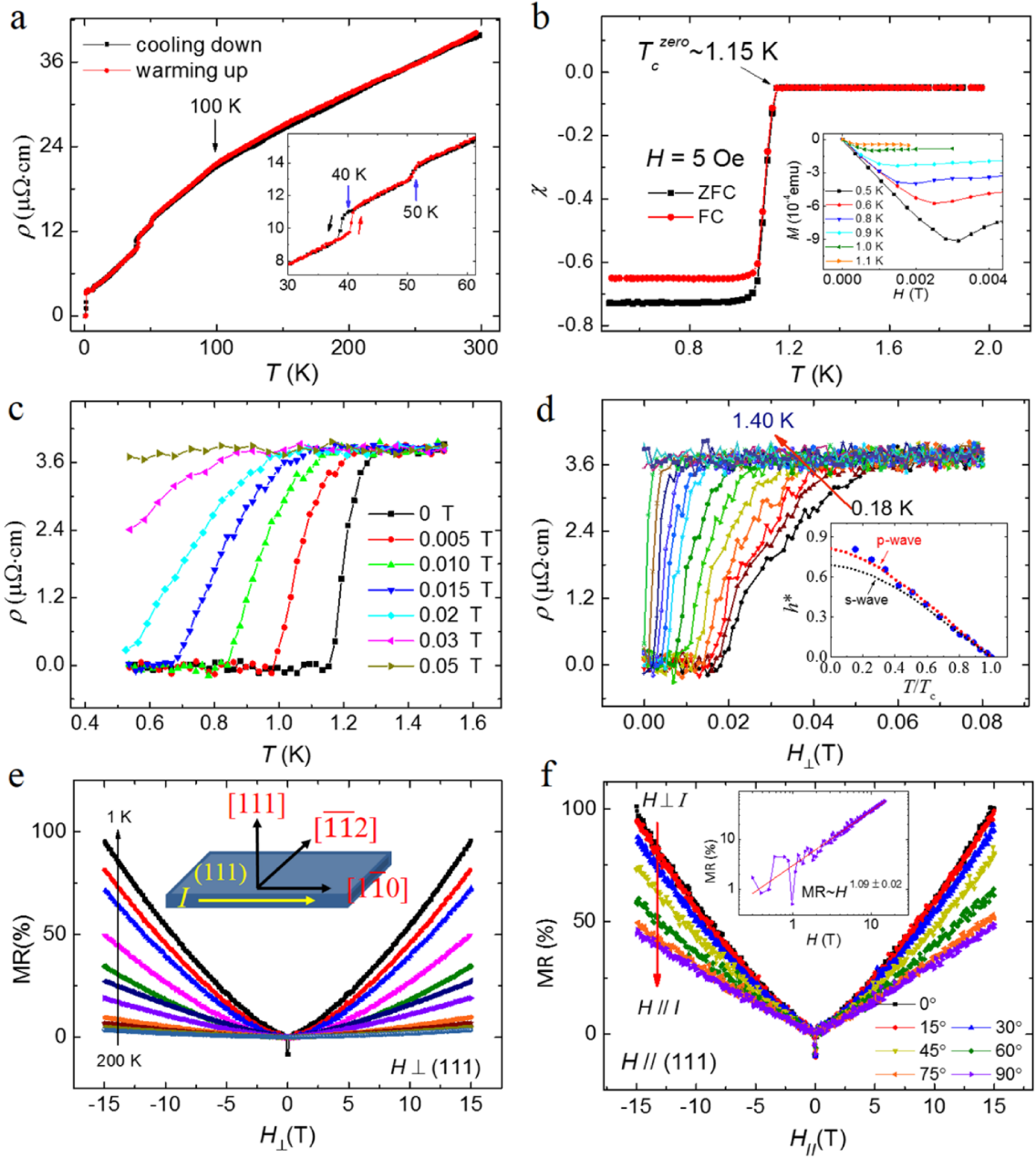

**Figure 2 Sample measurement of Au$_2$Pb.** (**a**) $\rho(T)$ curves of Au$_2$Pb single crystal. Inset: close-up of the same data from 30 K to 60 K. (**b**) Temperature dependence of χ, shows the Meissner effect: sharp diamagnetic drops at 1.15 K. The inset presents low-field $M(H)$ curves at various temperatures from 0.5 to 1.1 K. (**c**) $\rho(T)$ characteristics at various $H_\perp$ up to 0.05 T. The $\rho(T)$ obtained in zero field shows $T_c^{onset}$= 1.3 K, $T_c^{zero}$= 1.18 K. (**d**) Magnetoresistance of Au$_2$Pb crystal at various temperatures under $H_\perp$. Inset: Normalized upper critical field $h^*=H_{c2}/T_c dH_{c2}/dT|_{T=Tc}$ as a function of normalized temperature $t = T/T_c$, with the red dashed line indicating the expectation for a polar p-wave state.[19] The black dashed line indicates the WHH theory for s-wave superconductor.[18] (**e**) MR at temperatures 1 K, 5 K, 10 K, 20 K, 30 K, 50 K, 80 K, 100 K, 120 K, 150 K, 200 K measured in $H_\perp$. (**f**) Parallel magnetic field $H_{//}$ dependent resistivity at $T$ = 2.5 K with $H_{//}$ varying from the 0° to 90°. Inset: MR($H$) curves at 2.5 K for $H // I$ plotted on a logarithmic scale. The linear fitting gives the slope ∼ 1.09 ± 0.02.

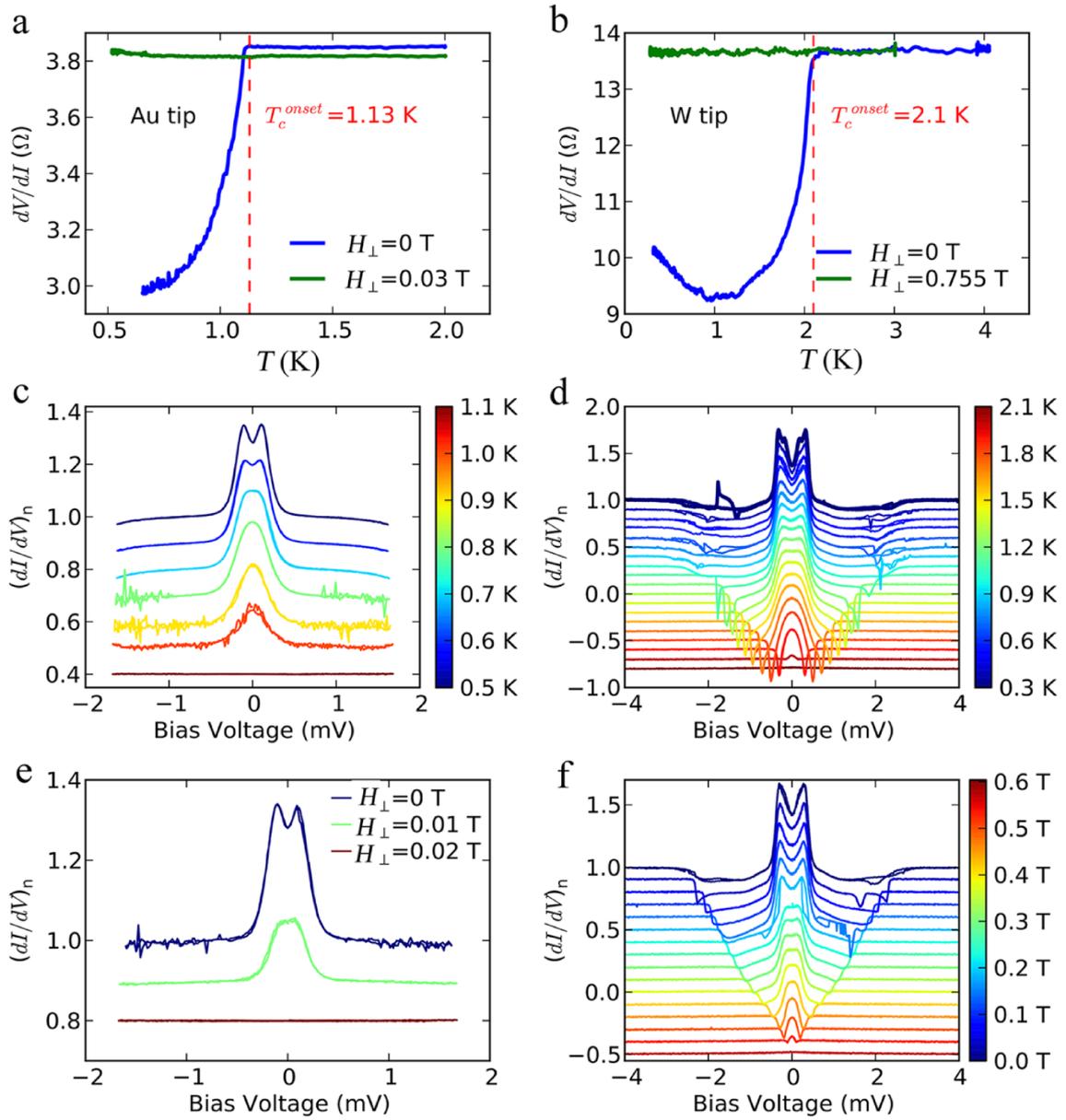

**Figure 3 PC results with the gold tip and the W tip.** (**a**)-(**b**) The zero bias PC resistance as a function of temperature for the gold tip and the W tip respectively; (**c**)-(**d**) the PCS at different *T*; (**e**)-(**f**) the PCS under different *H* at 0.5 K; curves shifted in (**c**)-(**f**) for clarity.

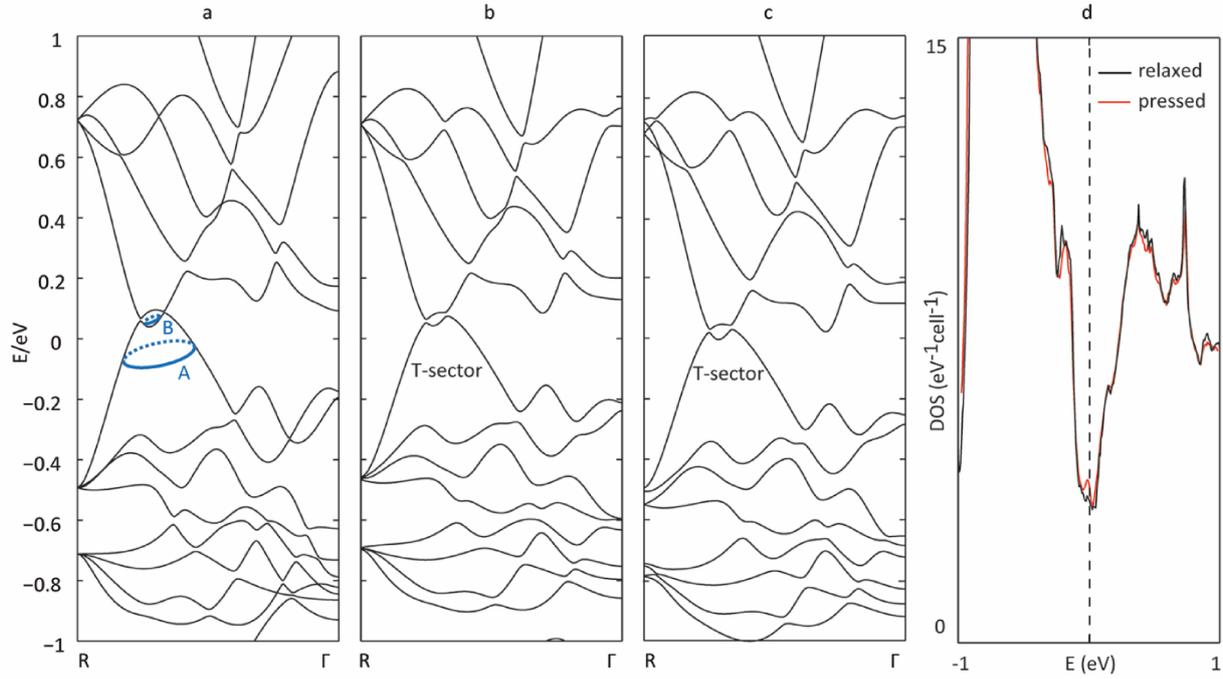

**Figure 4 The band structures of Au$_2$Pb along the Brillouin zone path R-Γ, which undergoes the most obvious change under 1% uniaxial compression.** (**a**) Calculation was done with relaxed structure and without spin-orbital coupling. The loop A and loop B correspond to the outer loop and inner loop in Fig. 5(a), respectively. (**b**) Calculation was done with relaxed structure and with spin-orbit coupling. The result is quite similar to (**a**). (**c**) Calculation was done with the structure under 1% uniaxial compression and with spin-orbit coupling. The point R is still defined by R = (b$_1$+b$_2$+b$_3$)/2, as in the case of relaxed structure. Note that the conduction band and the valence band shift to each other as compared to (**b**). This Brillouin zone region noticeably responsive to external pressure is referred to as the T-sector hereafter, and will be focused on in the investigations of orbital texture. (**d**) Comparison of the density of states of the relaxed structure (black curve) and the compressed structure (red curve). Spin-orbit coupling is included. The density of states near the Fermi level (shifted to 0 eV) is enhanced under the applied external pressure.

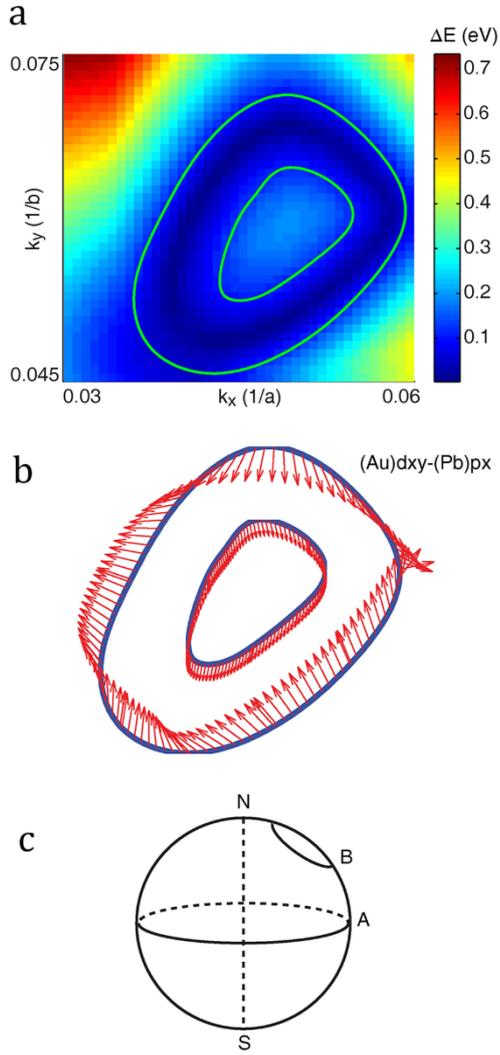

**Figure 5 Orbital texture of Au$_2$Pb.** (**a**) A small window of the cross section k$_z$= 0.69π (see Fig. 6(c)) of the Brillouin zone, which is near a T-sector. The background color indicates the energy difference of the conduction band and the valence band (without SOC). Note that there is a dark blue curve on which the conduction band and the valence band are (nearly) degenerate, so the orbital texture will suffer from discontinuity while going across this curve. Based on this consideration, we choose the two green loops, which avoid the degeneracy curve, to calculate the orbital textures. (**b**) The orbital texture of the phase difference of an Au-d$_{xy}$ orbital and a Pb-p$_x$ orbital on the valence band. The direction of the arrows indicates the phase difference of different orbital components of the corresponding wave function. The winding number of the orbital texture on the inner loop is 0, and the winding number of the orbital texture on the outer loop is -1. (**c**) The two closed loops partially demonstrate the orbital texture of the effective Hamiltonian (1). The winding number is -1 for the orbital texture on the loop A and is 0 for the loop B, corresponding to the cases of the outer and inner loops in (**b**), respectively.

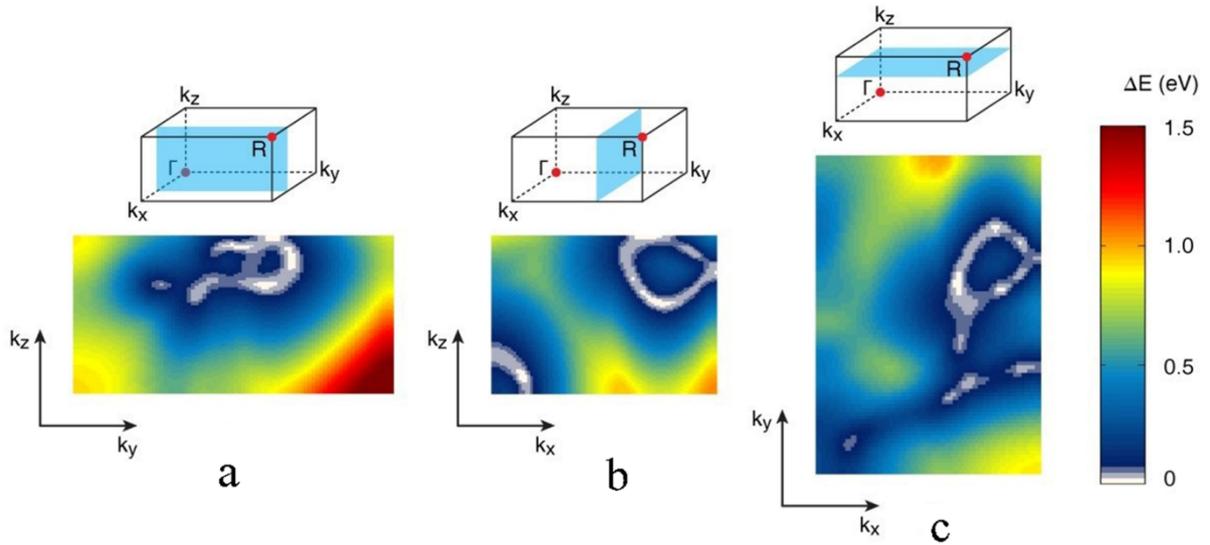

**Figure 6 Energy difference of the lowest conduction band and the top valence band in the Brillouin zone slices of** (**a**) $k_x=0.65\ \pi$, (**b**) $k_y=0.65\ \pi$, (**c**) $k_z=0.69\ \pi$. The top panels show the positions of the corresponding slices within the 1/8 Brillouin zone, $[0, \pi] \times [0, \pi] \times [0, \pi]$. Note that the background of Fig. S8a is a small region in the 2 o'clock direction of the panel (**c**). There are prominent white curves in all these panels, on which the lowest conduction band and the top valence band is nearly degenerate; i.e., these white curves indicate the existence of (quasi) nodal surfaces around the Fermi level.